\documentclass[prl,aps,preprint,showpacs]{revtex4}
\begin{document}
\title{Formation of Wigner crystals  in conducting polymer nanowires}
\author{Atikur Rahman}
\author{Milan K. Sanyal}
\affiliation{{\it Surface Physics Division, Saha Institute of
Nuclear Physics, 1/AF Bidhannagar, Kolkata 700 064, India.}}
\date{\today}
\begin{abstract}
The search for theoretically predicted Wigner crystal in
one-dimensional (1D) wires of structurally disordered materials
exhibiting properties of charge-density-waves have remained
unsuccessful. Based on the results of a low temperature conductivity
study we report here formation of such 1D Wigner crystal (1DWC) in
polypyrrole nanowires having low electron densities. The
current-voltage characteristics of all the nanowires show a 'gap'
that decreases rapidly as the temperature is increased - confirming
the existence of long-range electron-electron interaction in the
nanowires. The measured current show power-law dependence on voltage
and temperature as expected in 1DWC. A switching transition to
highly conducting state has been observed above a threshold voltage,
which can be tuned by changing the diameters of the nanowires and
the temperature. Negative differential resistance and enhancement of
noise has been observed above the threshold as expected.
\end{abstract}
\pacs{73.20.Mf, 73.20.Qt, 73.61.Ph, 73.63.-b}
\maketitle
Possibilities of applications in nanotechnology has triggered
extensive research activities to understand charge-conduction in
quasi-1D conductors like nanowires and nanotubes where
electron-electron interaction (EEI) plays a dominant role. In 1934
Eugene Wigner considered the effect of long-range EEI in metals and
predicted in a seminal paper \cite {wc0} possibility of formation of
periodic spatial structures of electrons for very low electron
density materials. It has been predicted theoretically that one
dimensional Wigner crystal (1DWC) \cite{wc1} formation exhibiting
the characteristic of a charge density wave (CDW) may occur in
nanowires of even structurally disordered materials \cite{dipole}.
Wigner crystal (WC) phase has been observed experimentally in
two-dimensional electron system under intense magnetic field
\cite{goldman, andrei}, in surface-state electrons of superfluid
helium \cite{jiang}, in quasi-1D organic charge transfer salt
\cite{hiraki} and recently in inorganic chain compounds
\cite{horsch}. Although interesting phases like L\"{u}ttinger liquid
(LL) \cite{LL}, which arise due to short-range EEI, has been
observed in several one dimensional (1D) systems \cite{nt1,nt2, ecb,
nw1}, predicted formation of 1DWC  in structurally disordered
materials \cite{dipole} has not been observed experimentally, to the
best of our knowledge. Among various low-dimensional systems
conducting polymer nanowires are easy-to-form quasi-1D systems to
study EEI as one can tune the carrier concentration of a polymer
over several orders of magnitudes by controlling doping
concentration and WC formation has been predicted in conducting
polymers \cite{rmpheeger,dipole}.

Here we report results of low temperature study of polypyrrole
nanowires having diameters from 30 to 450 nm. Highly non-linear
current-voltage (I-V) characteristics having $I\propto V^{1+\beta}$
at high V and $I\propto T^{\alpha}V$ at low V was observed. The
exponent $\beta$ reduces from high value ($\sim$5 -7) with
increasing temperature and does not take equal value to $\alpha$, in
general. A large 'gap' voltage
($V_{G}$), above which the conductance increase substantially, was
observed in the low temperature I-V characteristics. The gap was
found to decrease rapidly with increasing temperature and with
increasing diameter. The I-V characteristics also show a switching
transition to a highly conducting state above a certain threshold
voltage $V_{Th} (>V_G)$. The threshold field ($E_{Th}$)
corresponding to $V_{Th}$ was found to depend on the diameter (d) as
$E_{Th}\propto d^{-4/3}$. Current-driven I-V characteristics show
negative differential resistance (NDR) at low temperature. An
increase in 'noise' has been observed in the switched (highly
conducting) state. All these experimental results give evidence in
favor of pinned Wigner crystal formation in conducting polymer
nanowires, as predicted theoretically \cite{dipole}.

Conducting polymer nanowires are quasi-1D systems composed of
aligned polymer chains where charge carriers are created by doping.
We used membrane based synthesis technique for growing polymer
nanowires following same preparation technique describe earlier
\cite{prb1}. The average doping concentration (obtained after
initial doping gradient within few hundred nanometer) can be
systematically reduced by lowering the diameters of the nanowires
\cite{prb1}. The actual diameter of nanowires were characterized
from SEM micrograph and were found to have an average diameter of
30, 50, 70, 110, 350 and 450 nm. Gold electrodes (2mm diameter) were
sputter-deposited on both sides of the membrane to establish
parallel connection with $\sim 10^7$ nanowires (pore density of the
membrane $\sim 10^9/$ cm$^2$). The electrical measurements were
carried out in a Oxford cryostat under liquid Helium environment
using Keithley 2400 source-meter and 6517A electro-meter in two
probe configuration (refer inset of Fig. 1(a)) over the temperature
range of 1.7 to 300 K. The consistency of the results were checked
by changing the contact material, scan-speed of the bias voltage
(current). Differential conductance ($dI/dV$) were obtained by
numerically differentiating the I-V curves and verified by Lock-in
measurements. In Fig. 1(a) we have shown $dI/dV$ vs. $V$ plots for
30 nm nanowires taken at various temperatures.

The conductance vs. voltage and $dI/dV$ vs. voltage data of various
nanowires (refer Fig. 1(a) and (b)) show existence of 'gap' voltage
($V_G$). We have measured the value of $V_G$ by noting the change of
slope from the conductance vs. voltage data for different diameter
nanowires as shown in Fig. 1(b). $V_{G}$ was found to be inversely
proportional to the diameter ($d$) of the nanowires (refer inset of
Fig. 1(b)). It is to be noted that zero-bias current and hence the
conductance below the $V_G$ increases with temperature and diameter
of nanowires. The 'gap' was found to reduce rapidly with increase in
temperature and vanishes at relatively high temperature that depends
on the wire diameter.

Strong temperature dependence of 'gap' is a signature of
electron-electron interaction \cite{boris, cb1}, and has been
observed earlier in chains of graphitized carbon nanoparticle
\cite{cb2}. Temperature and diameter dependence of 'gap' in our
nanowires can only be explained by considering 'pinned' collective
state \cite{dipole} where pining strength increases with decreasing
diameter. This is consistent with the fact that observed voltage
required for switching transition also increases with decreasing
diameter. In a previous study existence of collective behavior has
been predicted in these nanowires as a non-Curie type temperature
dependence of static dielectric constant \cite{prb1} was observed.
Here dipoles (formed by dopant counter ion) interact among each
other via Coulomb interaction and produce a collective pinning
\cite{dipole}. The observed characteristics of $V_G$ here is
consistent with Wigner crystal formation  which is pinned by the
impurity \cite{dipole, andrei,jiang}.

Above the voltage $V_{G}$, I-V characteristics of all nanowires show
power law behavior - a known characteristics of 1D conductors. In
contrast to conventional three-dimensional (3D) materials, 1D
conductors exhibit fascinating transport properties due to the power
law dependence of tunneling density of states, which can be
parameterized as $dI/dV \propto T^\alpha$ and $\propto$ V$^\beta$
for low bias ($V \ll k_BT/e$) and high bias ($V \gg k_BT/e$)
conditions respectively \cite{LL}. In Fig. 2(a) we have shown a
representative data that show $R\propto T^{-\alpha}$ behavior. In
this figure we have also plotted the same data as $\log(R)$ vs.
$T^{-1/4}$ to show that variable range hopping (VRH) can also give
reasonable fit. We found that for low bias data VRH give better fit
\cite{prb1} but power law gives better fit for the data taken with
higher bias, as discussed earlier \cite{artmenko}. In the inset of Fig. 2(a) we have
plotted resistivity data (taken with bias 1V) against temperature in
log-log scale for 30nm diameter nanowire and fitted it to show that
power law dependence ($R\propto T^{-\alpha}$, with $\alpha = 5.2$)
remains above 30K. Due to the presence of $V_G$, below 30K the
resistance for this bias becomes nearly temperature independent as
observed in other polymer nanowires \cite{alshn}. For 30 nm diameter
nanowire taking $\alpha = 5.2$ and plotting $I/T^{1+\alpha}$ versus
$eV/(k_B T)$, various I-V curves of different temperatures collapse
on a master curve (Fig. 2(b)). Clean LL state predicts $\alpha =
\beta$ and scaling of I-V curves of different temperatures to such a
master curve \cite{nt1,nt2, arx}. But for the higher diameter
nanowires the I-V characteristics of different temperatures could
not be collapsed to a single master curve. Presence of $V_G$,
absence of single master curve and unequal exponents (discussed
below) $\alpha$ and $\beta$ in these wires show that LL theory is
not applicable.

The power law behavior is a characteristic feature of 1D transport
\cite{nt1,nt2,nw1,LL,wc1,ecb,gia,jeon} and has also been observed in
nanowires of conventional CDW materials \cite{slot2}. From the I-V
data we get $\beta= 5.6$ at T = 3 K for 30 nm diameter nanowire
(refer upper inset of Fig. 2(b)) and for other diameter nanowires
this value goes up to 7.2 at 3K (refer lower inset of Fig. 2(b)).
Similar power law behavior have been observed in other 1D systems
like carbon nanotubes ($\alpha,\beta \sim 0.36$) \cite{nt1} and
nanowires of InSb ($\alpha \sim 2-7, \beta \sim 2-6$) \cite{nw1},
polymer ($\alpha \sim 2.2-7.2, \beta \sim 2-5.7$) \cite{alshn},
NbSe$_3$ ($\alpha \sim 1-3, \beta \sim 1.7-2.7$) \cite{slot2} and
MoSe ($\alpha \sim 0.6-6.6, \beta \sim 0.32-4.9$) \cite{MoSe} .
Although we obtained $\alpha \simeq \beta$ for most of the nanowires
at low temperature but the $\beta$ values were found to decrease
with increasing temperature and with decreasing diameter (refer
lower inset of Fig. 2(b)) - as a result $\alpha$ remains unequal to
$\beta$, in general for our wires.

For a quasi-1D system strength of Coulomb correlation is defined as
$r_s=a/(2a_B)$, where $a$ is the average distance between electrons
and $a_{B}$ is the effective Bohr radius. Low doped Polymer
nanowires with quasi-1D nature and low electron density ($r_s\gg 1$)
are a potential candidate to form Wigner crystal that can exhibit
characteristics of a charge density wave state \cite{wc1,rmpheeger,
dipole,alshn,ppycdw}. For weakly pinned Wigner crystals tunneling
density of states show a power law behavior with the applied bias
\cite{gia} and the exponent ranges from $\sim 3$ to 6
\cite{hclee,jeon}. It has been shown \cite{jeon} that for 1DWC with
increasing pinning strength $\beta$ should decrease. The variation
of $V_{\mathrm G}$ with d (refer inset of 1(b)) clearly indicates
that pinning strength increases \cite{cb1} with decreasing diameter.
Hence our observation of reduction in $\beta$ value with decreasing
diameter of nanowires is consistent with 1DWC model. Moreover higher
values of the exponents observed here also indicate that 1DWC has
formed in our nanowires.

At low temperature all the nanowires show a switching transition to
a highly conducting state above a certain threshold voltage $V_{Th}
(>V_{G})$. We did not observe any switching transition when the bias
voltage is kept $\sim$ 1 mV below $V_{Th}$ for long time. The sharp
threshold indicates a collective phenomena \cite{sw2,sw1,gruner} and
the transition is not due to field heating. In Fig. 3(a) and in its
inset we have shown representative switching transitions for the
nanowires measured at 2.5 K. All the nanowires showed hysteresis in
the switching transition which is independent of the bias scan speed
(thus removes any possibility of capacitive effect). The nanowires
could switch back to the low conducting state only when the applied
voltage is reduced to a value $V_{Re}$
($|V_{Th}|>|V_{Re}|>|V_{G}|$). With decreasing temperature the
hysteresis, defined as ($E_{Th} - E_{Re})/E_{Re}$ ( $E_{Re}$ is the
field corresponding to $V_{Re}$), increases (refer Fig.3(b))- this
behavior is consistent with that observed in switching CDW
\cite{sw2,gruner,sw1,twoth}. It has been shown theoretically that
formation of 1DWC is equivalent to have $4k_{F}$ CDW in a system
\cite{dipole,wc1}. The sliding state of this pinned CDW can explain
the field induced switching transition observed here. Presence of
$V_{G}$ and $V_{Th}$ is also consistent with two threshold observed
in semiconducting CDW systems \cite{twoth}.

Depending upon the pinning strength a pinned CDW become
non-conducting below a certain threshold field. When an applied DC
field is strong enough to overcome the pinning energy the CDW depins
and sliding motion starts giving rise to a switching transition
\cite{sw2,gruner,sw1,twoth}. When CDW is confined in two direction
as in the nanowires, phase deformations occur along the length of
the nanowires. In this situation pinning of CDW is one dimensional
and the threshold field is expected to be proportional to $d^{-4/3}$
\cite{slot1}. The plot of $E_{Th}$ vs. $d$ shown in Fig. 3(c)
confirms this dependence in our nanowires. The change in $E_{Th}$
with $d$ is obviously not due to surface pinning as that would have
given us $E_{Th}\propto d^{-1}$. It has been also observed
previously in a study of NbSe$_3$ samples that surface pinning can
be excluded for highly resistive samples ($R/L>1\Omega/\mu$m)
\cite{slot1}. We measured current-driven I-V characteristics to
investigate the nature of switched state in these nanowires. The
measurements (Fig. 3(d) and its upper inset ) exhibit negative
differential resistance (NDR). This type of behavior has been
observed in sliding CDW state \cite{twoth,hall} and is expected in
1DWC. The comparison between voltage driven and current driven I-V
characteristics has been shown in Fig. 3(d) and this shows the
uniqueness of threshold field for both type of measurements. We have
also observed a large fluctuation in the measured voltage in the
switched state (refer lower inset of Fig. 3(d)), which gives strong
evidence of sliding motion above $E_{Th}$ \cite{twoth}, similar
increase in 'noise' has been observed previously in two dimensional
Wigner crystal \cite{goldman}.

The results presented here show that the I-V characteristics of all
the nanowires have three distinct regions (refer Fig. 3(a)). Below
$V_G$ current is very small and that increases with temperature and
diameter. Between $V_G$ and $V_{Th}$ power-law characteristics of 1D
transport is observed. Above $V_{Th}$ switching transition is
observed that exhibit hysteresis, $d^{-4/3}$ scaling, NDR and
'noise' enhancement. All these findings confirm the formation of
1DWC which is expected to exhibit characteristics of CDW, in our
nanowire. Although power law dependent I-V characteristics can be
explained by other models like L\"{u}ttinger liquid \cite{LL} and
environmental Coulomb blockade (ECB) \cite{glazman}, these theories
cannot explain the observed switching transition and related
phenomena reported here. Moreover, LL theory is clearly inconsistent
with observed $V_G$ and reduction of $\beta$ with decreasing
diameter (refer inset of Fig. 2(b)). In addition we could not get
expected collapse to a master curve \cite{arx} for higher diameter
nanowires as observed in 30nm wires (refer Fig. 2(b)). ECB theory
can account for $V_G$ but it predicts an increase in $\beta$ with
increasing environmental impedance but in our case though resistance
of the nanowires decreases with increasing diameter - $\beta$
increases. In conclusion all our experimental findings show that one
dimensional Wigner crystal has been formed in nanowires of a
structurally disordered material like conducting polymer.

Authors are grateful to R. Gangopadhayy and A. De for their help in
sample preparation.

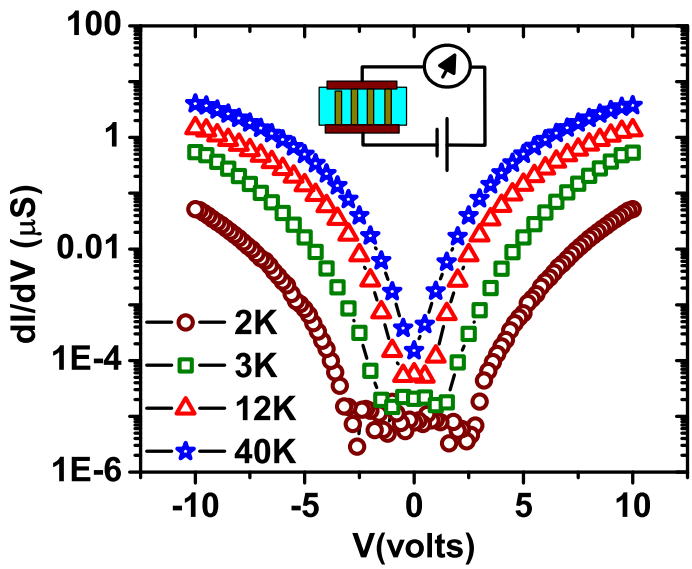
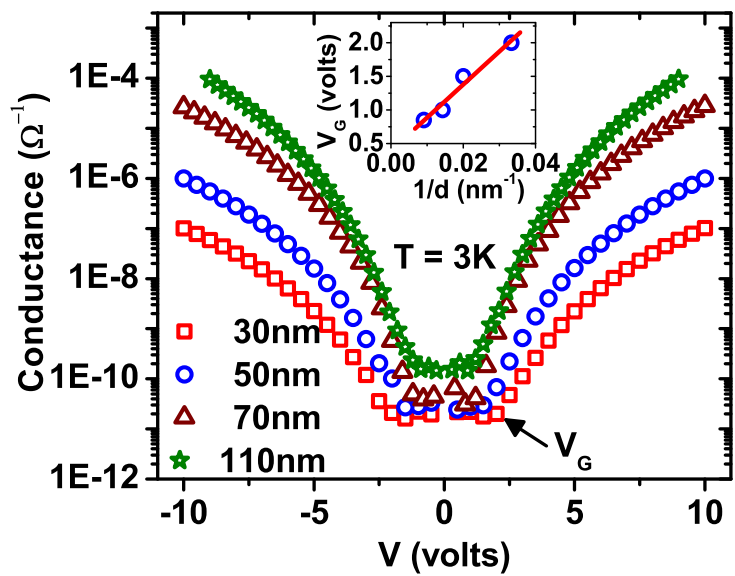
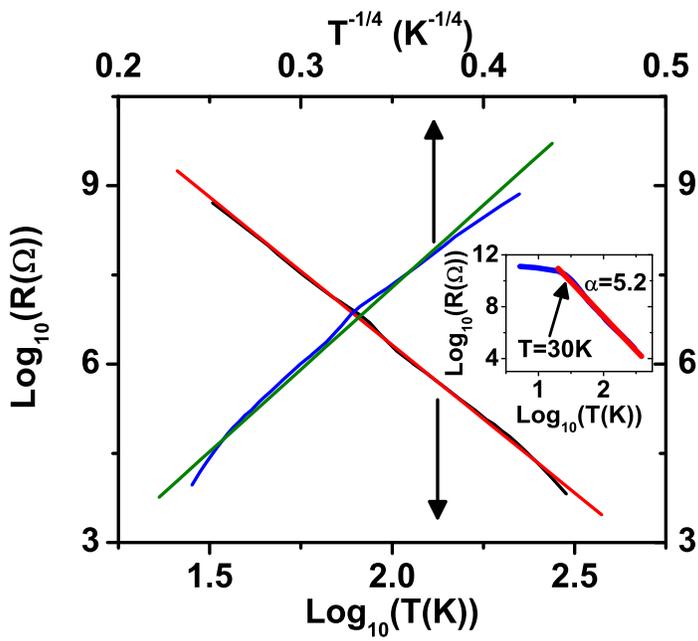
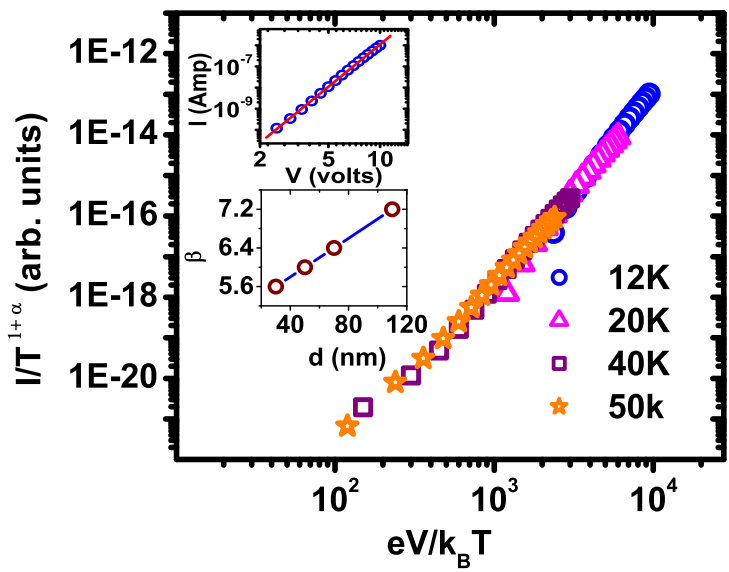
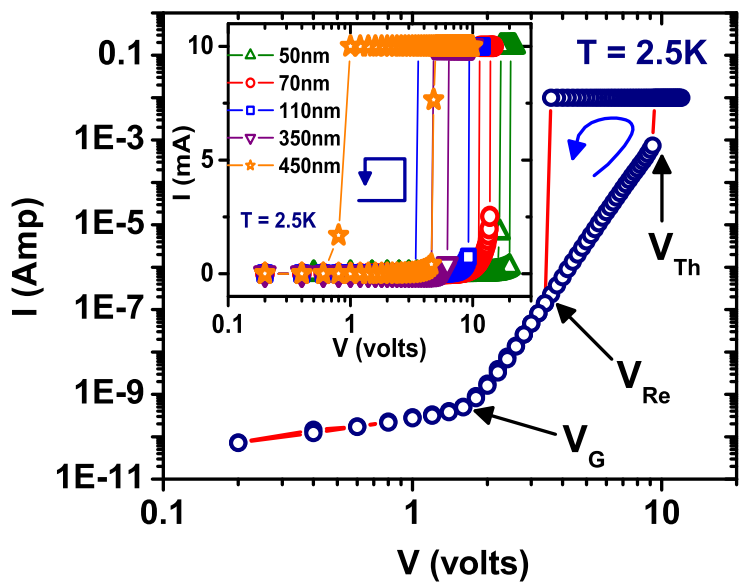
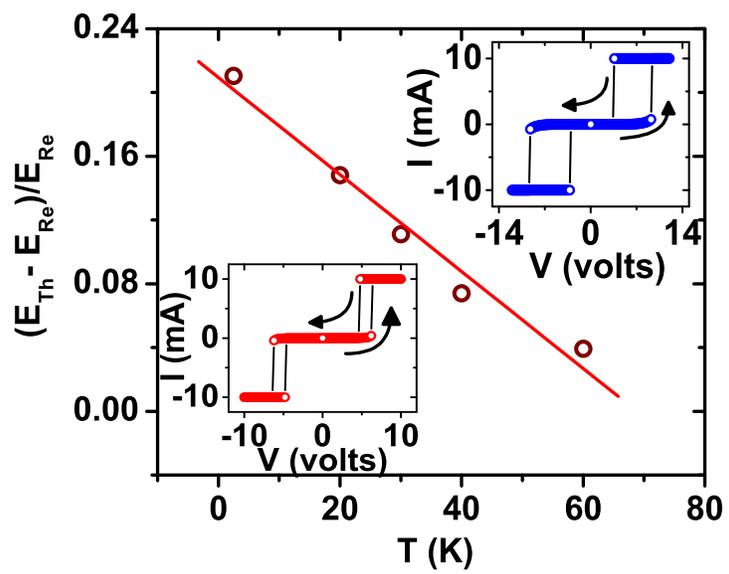
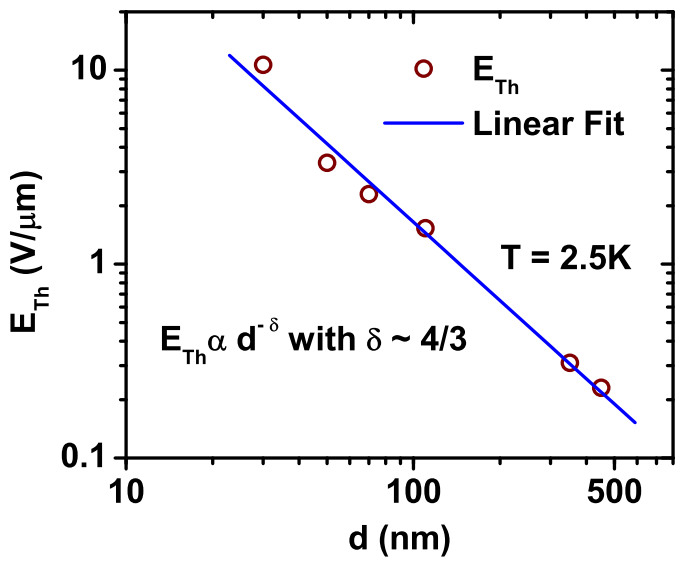
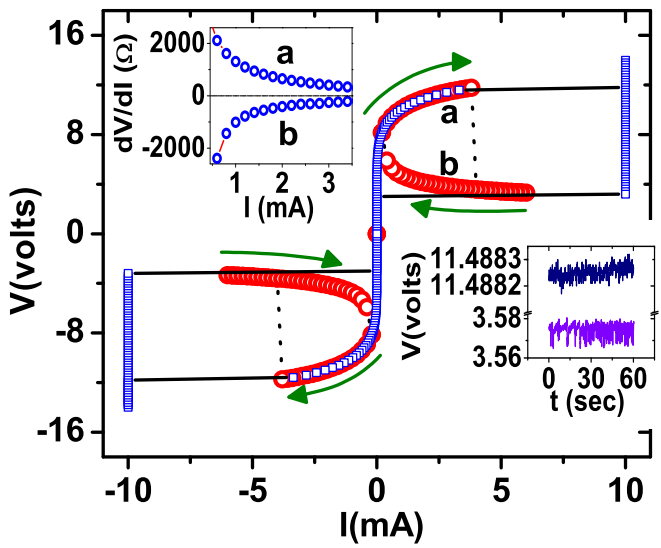
\end{document}